# Statistical mechanics of high-density bond percolation

P. N. Timonin

*Physics Research Institute, Southern Federal University, 194 Stachki Avenue, Rostov-on-Don, 344090 Russia,* e-mail: pntim@live.ru

*High-density (HD) percolation describes the percolation of specific $\kappa$ - clusters, which are the compact sets of sites each connected to κ nearest filled sites at least. It takes place in the classical patterns of independently distributed sites or bonds in which the ordinary percolation transition also exists. Hence, the study of series of $\kappa$ - type HD percolations amounts to the description of classical clusters' structure for which $\kappa$ - clusters constitute κ – cores nested one into another. Such data are needed for description of a number of physical, biological and information properties of complex systems on random lattices, graphs and networks. They range from magnetic properties of semiconductor alloys to anomalies in supercooled water and clustering in biological and social networks. Here we present the statistical mechanics approach to study HD bond percolation on arbitrary graph. It is shown that generating function for $\kappa$ - clusters' size distribution can be obtained from the partition function of specific q – state Potts-Ising model in $q \to 1$ limit. Using this approach we find exact $\kappa$ - clusters' size distributions for Bethe lattice and Erdos – Renyi graph. The application of the method to Euclidean lattices is also discussed.*

## I. Introduction

A number of physical properties of disordered materials depend on the structure of the clusters they contain. Among them are the ferromagnetism of dilute semiconductors [1], the catalytic ability of random films [2], electrolytic dissolution of binary alloys [3], diffusion in a crowded environment [4] and many others. Phase transitions in such materials often result from percolation transitions – emergence of giant cluster of specific sort, relevant for specific property of the material. Mechanisms of some phase transforms in random media such as stainless steel corrosion [5] and anomalies in super cooled $H_2O$ and $D_2O$ [6] are possibly related to the emergence of various compact tightly bound clusters. High-density (HD) percolation was just introduced for description of percolation via clusters of varying compactness, which are the set of sites each connected to κ nearest filled sites at least ( $\kappa$ - clusters for short) [6, 7, 8]. Contrary to many other models of correlated percolation [9] original HD percolation [6, 7, 8] takes place in the classical uncorrelated patterns of randomly and independently distributed sites or bonds which feature also the classical percolation transition, that is, the transition with κ = 0, 1. Apparently, HD κ – clusters constitute what can be termed as κ – cores of conventional clusters (see Fig. 1) while HD giant κ – clusters are the κ – cores of usual giant component of classical site or bond percolation. Accordingly, it is shown in Refs. [7, 8] that in classical site percolation on Bethe lattice such giant κ – cores emerge at concentrations above the classical percolation threshold $p_\kappa > p_c = p_1 = p_0$, κ > 1 and $p_{\kappa+1} > p_\kappa$. Thus, the sequence of HD percolation transitions at $p_\kappa$ manifests the appearance of more and more connected infinite κ – cores (one nested to another) in the usual percolation cluster. This is common feature of phase transitions with nonlocal order parameters – they allow for the multitude of other nonlocal order parameters and cascades of corresponding transitions [10].

The conventional way to study the critical properties of HD percolation transitions is to find the clusters' size distribution for them. In the implicit form, this has been done for HD site percolation on Bethe lattice using the theory of random walks [7]. For some 2D and 3D lattices this HD distributions are found numerically via enumeration of clusters of κ or more coordinated sites in random bonds [6] and sites [11, 12] patterns. Yet such numerics is very time-consuming for sufficiently large samples. So one may try to diminish the computational problems via casting HD percolation into statistical mechanics framework as has been done for classical bond percolation [13, 14]. This may help to diminish the numerical efforts due to a number of methods developed for partition function calculation such as transfer matrix technique in conjunction with renormalization group and Monte Carlo simulations.

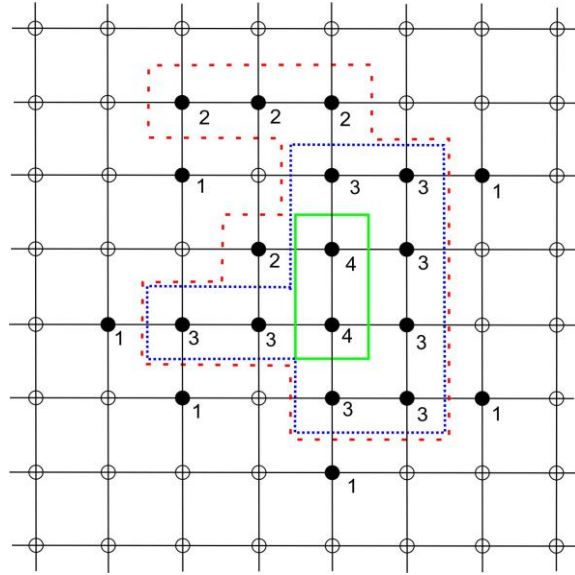

Fig. 1. *The example of the decomposition of the ordinary cluster on square lattice (black circles) into κ – clusters. The numbers at cluster's sites denote their coordination numbers. Full line encircles 4 – cluster, dotted line encircles 3 – cluster and dashed line encircles 2 – cluster.*

We can expect that such high-precision numerics would establish the reliable values of transition points, critical indexes, their scaling relations and universality classes of HD percolation transitions on various graphs and Euclidean lattices. The statistical mechanics approach could also give some exact results for various hierarchical lattices and deterministic fractals through simple algebraic derivation, i.e. without resorting to the probability theory constructions.

Here we show how generating function of $\kappa$ - clusters' size distribution for HD bond percolation can be obtained on arbitrary graph from the $q \to 1$ limit of partition function of specific q-state Potts-Ising model using the ideas of Refs. [13, 14]. In Section II we describe the general formalism, in Section III we apply it to obtain the exact clusters' size distribution for HD bond percolation on Bethe lattice and Erdos – Renyi graph, and in Section IV and Appendix we discuss its application to the Euclidean lattices

## II. General formalism.

The present method is based on the Fortuin-Kasteleyn's cluster representation of Potts model [13] and its implementation for classical bond percolation [14].

Consider a graph with *N* sites and set of edges *E*. To describe the bond configurations on it we assign to each edge the variable $n_{i,j} = 0, 1$ with 0/1 corresponding to the absence/presence of bond. For each bond present with probability *p* we have the probability of a general configuration

$$W(\mathbf{n}) = \prod_{i,j \in E} \rho(n_{i,j}) , \quad \rho(n_{i,j}) = (1-p)(1-n_{i,j}) + pn_{i,j} = 1 - p + n_{i,j}(2p-1)$$

Our task is to discern in each bond configuration the clusters of sites having at least $\kappa$ bonds attached. As in original mapping of classical bond percolation problem onto *q* - state Potts model [13, 14] to do this we should provide the bonds of these $\kappa$ - clusters with the factor $\delta(\sigma_i, \sigma_j)$ (Kronecker delta) making all Potts variables $\sigma = \{0, 1, ..., q-1\}$ in a $\kappa$-cluster equal. This job is done by putting to each edge the factor

$$g^{(\kappa)}_{\sigma_i,\sigma_j}\left(\mathbf{n},k_i,k_j\right)=1+n_{i,j}\vartheta\left(k_i-\kappa\right)\vartheta\left(k_j-\kappa\right)\left[\delta\left(\sigma_i,\sigma_j\right)-1\right] \quad (1)$$

Here $\vartheta(x)$ is Heaviside step function, $\vartheta(0)=1$, and $k_m = \sum_{\langle l,m\rangle\in E} n_{m,l}$ is the number of bonds attached to the $m$-th site in a given configuration. So when bond is present on the $\langle i,j\rangle$ edge and the number of bonds attached to $i$-th and $j$-th sites exceeds $\kappa-1$ then $g^{(\kappa)}_{\sigma_i,\sigma_j}\left(\mathbf{n},k_i,k_j\right)=\delta\left(\sigma_i,\sigma_j\right)$, otherwise $g^{(\kappa)}_{\sigma_i,\sigma_j}\left(\mathbf{n},k_i,k_j\right)=1$.

Also we need the fields at each site [14]

$$h_{\sigma_i}=h\left[\delta(\sigma_i,0)-1\right] \quad (2)$$

to count the number of sites in the $\kappa$-clusters. Thus, we have the partition function, which contains essential data for $\kappa$-type HD bond percolation

$$Z_\kappa(p,h,q) = Tr_{\boldsymbol{\sigma},\mathbf{n}}W(\mathbf{n})\prod_{\langle i,j\rangle\in E}g^{(\kappa)}_{\sigma_i,\sigma_j}\left(\mathbf{n},k_i,k_j\right)\prod_{i=1}^{N}\delta\left(k_i,\sum_{\langle l,i\rangle\in E}n_{i,l}\right)\exp h_{\sigma_i} \equiv Tr_{\boldsymbol{\sigma},\mathbf{n}}\exp(-\mathcal{H}_\kappa) \quad (3)$$

Indeed, in the cluster representation using the equality of Potts variables $\boldsymbol{\sigma}$ on sites belonging to $\kappa$-clusters we get

$$Z_\kappa(p,h,q) = \sum_{C} p^{B(C)}(1-p)^{|E|-B(C)}\prod_{\kappa-clusters}\left[1+(q-1)e^{hs_{cl}}\right] = \left\langle\prod_{s=1}\left[1+(q-1)e^{hs}\right]^{N^{(\kappa)}_s(C)}\right\rangle_C \quad (4)$$

where $B(C)$ is the number of bonds in configuration $C$, $N^{(\kappa)}_s(C)$ is the number of $\kappa$-clusters with $s$ sites in it and $\langle...\rangle_C$ means the average over bond configurations

$$\langle A(C)\rangle_C = \sum_{C} p^{B(C)}(1-p)^{|E|-B(C)}A(C).$$

Hence, at $q\to 1$

$$Z_\kappa(p,h,q) \approx 1+(q-1)NG_\kappa(p,h) \quad (5)$$

$G_\kappa(p,h)$ being the generating function for $\kappa$-clusters' size distribution

$$G_\kappa(p,h)=\sum_{s=1}v^{(\kappa)}_s e^{hs}, \quad v^{(\kappa)}_s = \left\langle\frac{N^{(\kappa)}_s(C)}{N}\right\rangle_C \quad (6)$$

The effective Hamiltonian $\mathcal{H}_\kappa$ defined in the last equality of Eq. 3 is linear in Potts interaction $\delta(\sigma_i,\sigma_j)$ and polynomial in Ising-like variables $n_{i,j}$. Thus we have specific Potts-Ising model with Potts spins at the sites and $n_{i,j}$ at the edges of a graph which describes essential properties of HD bond percolation at $q\to 1$.

### III. HD percolation on Bethe lattice

Bethe lattice with large coordination number $z$ can be sufficiently adequate approximation for highly coordinated Euclidean lattices as well as for Erdos-Renyi network when $z\to\infty$, and $pz\to c$ [15-17]. So we can test the consistency of described approach applying it to HD bond percolation on Bethe lattice the more so as this can give some analytical mean-field results for $\kappa$ - clusters' size distribution.

We use the standard method to obtain the density of thermodynamic potential $N^{-1}\ln Z_\kappa(p,h,q)$ of the model on Bethe lattice via the partial partition functions on Caley trees [18], which can be found from the recurrence relations between the trees of $l$-th and $l+1$-th levels. In the thermodynamic limit, $l \to \infty$, we find the stable stationary values for the parameters defining these partition functions and the density of thermodynamic potential. Then, according to (5), for $q \to 1$

$$\lim_{N\to\infty} N^{-1}\ln Z_\kappa(p,h,q) \approx (q-1)G_\kappa(p,h)$$

and we obtain the generating function $G_\kappa(p,h)$ for Bethe lattice.

First, we introduce the partial partition function of our model on $l$-level Caley tree $U_{\sigma,l}^{(\kappa)}(k,n)$ summed over all dynamic variables except the root ones, $\sigma$ being Potts variable of the root site, $k$ is the number of bonds attached to it and $n$ is the root edge variable. Recurrence relations for these quantities are

$$U_{\sigma,l+1}^{(\kappa)}(k,n) = \left[1-p+n(2p-1)\right] \sum_{\sigma',k',n_1,\ldots,n_{z-1}} \left\{ \begin{array}{l} 1+n\vartheta(k-\kappa)\vartheta(k'-\kappa)\left[\delta(\sigma,\sigma')-1\right]\exp h_{\sigma'} \times \\ \delta\left(k',n+\sum_{i=1}^{z-1} n_i\right)\prod_{i=1}^{z-1} U_{\sigma',l}^{(\kappa)}(k',n_i) \end{array} \right\} \quad (7)$$

Eq. 7 suggests the following form of $U_{\sigma,l}^{(\kappa)}(k,n)$:

$$U_{\sigma,l}^{(\kappa)}(k,n) = (1-n)a_l + n\left[b_l + \vartheta(k-\kappa)(c_{\sigma,l}-b_l)\right], \quad c_{\sigma,l} = c_{1,l} + \delta(\sigma,0)(c_{0,l}-c_{1,l}),$$

that is

$$U_{\sigma,l}^{(\kappa)}(k,0) = a_l, \qquad U_{\sigma,l}^{(\kappa)}(k,1)\Big|_{\kappa<k} = b_l, \qquad U_{\sigma,l}^{(\kappa)}(k,1)\Big|_{\kappa\geq k} = c_{\sigma,l} \quad (8)$$

From Eqs. 7, 8 we have

$$a_{l+1} = (1-p)\left\{ \begin{array}{l} \sum_{m=0}^{\kappa-1}\binom{z-1}{m}a_l^{z-1-m}b_l^m + \sum_{m=\kappa}^{z-1}\binom{z-1}{m}a_l^{z-1-m}c_{0,l}^m \\ +(q-1)e^h\left[\sum_{m=0}^{\kappa-1}\binom{z-1}{m}a_l^{z-1-m}b_l^m + \sum_{m=\kappa}^{z-1}\binom{z-1}{m}a_l^{z-1-m}c_{1,l}^m\right] \end{array} \right\} \quad (9)$$

$$b_{l+1} = p\left\{ \begin{array}{l} \sum_{m=0}^{\kappa-2}\binom{z-1}{m}a_l^{z-1-m}b_l^m + \sum_{m=\kappa-1}^{z-1}\binom{z-1}{m}a_l^{z-1-m}c_{0,l}^m \\ +(q-1)e^h\left[\sum_{m=0}^{\kappa-2}\binom{z-1}{m}a_l^{z-1-m}b_l^m + \sum_{m=\kappa-1}^{z-1}\binom{z-1}{m}a_l^{z-1-m}c_{1,l}^m\right] \end{array} \right\} \quad (10)$$

$$c_{1,l+1} = b_{l+1} - p\sum_{m=k-1}^{z-1}\binom{z-1}{m}a_l^{z-1-m}c_{0,l}^m + (2-q)e^h p\sum_{m=k-1}^{z-1}\binom{z-1}{m}a_l^{z-1-m}c_{1,l}^m \quad (11)$$

The combinatorial coefficients here obey the usual convention for $m>n$ or $m<0$.

$$\binom{n}{m} = 0$$

Parameters $a_l$, $b_l$ and $c_{\sigma,l}$ tend to some stable stationary points $a$, $b$ and $c_\sigma$ when $l \to \infty$. These stationary values define the density of thermodynamic potential of our model on Bethe lattice [18] as follows

$$\lim_{N \to \infty} \frac{2}{N} \ln Z_\kappa(h,p,q) = (2-z) \lim_{l \to \infty} \ln \sum_{\sigma, n_1, \ldots, n_z, m} \exp h_\sigma \delta\left(m, \sum_{i=1}^{z} n_i\right) \prod_{i=1}^{z} U_{\sigma,l}^{(\kappa)}(m, n_i) =$$

$$(2-z) \ln \sum_\sigma \exp h_\sigma \sum_{m=0}^{z} \binom{z}{m} \left[U_\sigma^{(\kappa)}(m,1)\right]^m \left[U_\sigma^{(\kappa)}(m,0)\right]^{z-m} =$$

$$(2-z) \ln \left\{ \sum_{m=0}^{\kappa-1} \binom{z}{m} a^{z-m} b^m + \sum_{m=\kappa}^{z} \binom{z}{m} a^{z-m} c_0^m + (q-1) e^h \left[\sum_{m=0}^{\kappa-1} \binom{z}{m} a^{z-m} b^m + \sum_{m=\kappa}^{z} \binom{z}{m} a^{z-m} c_0^m \right]\right\} = \quad (12)$$

$$(2-z) \ln \left\{ (a+b)^z + \sum_{m=\kappa}^{z} \binom{z}{m} a^{z-m} (c_0^m - b^m) + (q-1) e^h \left[\sum_{m=0}^{\kappa-1} \binom{z}{m} a^{z-m} b^m + \sum_{m=\kappa}^{z} \binom{z}{m} a^{z-m} c_0^m \right]\right\}$$

For $q \to 1$ we have from Eqs. 9-11

$$a = 1 - p + O(q-1), \quad b = p + O(q-1)$$

$$a + b \approx 1 + \frac{q-1}{2-z} e^h \left\{\begin{array}{l} 1 - p D_{\kappa-1}(p) - (1-p) D_\kappa(p) - (1-p)[D_{\kappa-1}(u) - D_\kappa(u)] + \\ D_{\kappa-1}(u)[1-(z-1)D_{\kappa-1}(p)] \end{array}\right\}, \quad (13)$$

$$c_0 \approx b - (q-1) e^h p D_{\kappa-1}(u)$$

Here

$$D_\kappa(u) \equiv \sum_{m=\kappa}^{z-1} \binom{z-1}{m} (1-p)^{z-1-m} u^m, \quad u \equiv c_1\big|_{q=1}. \quad (14)$$

Variable $u = c_1\big|_{q=1} = u(e^h, p, \kappa)$ is the solution to the equation (see Eq. 11)

$$u = u_\kappa(p) + e^h p D_{\kappa-1}(u), \quad u_\kappa(p) \equiv p[1 - D_{\kappa-1}(p)], \quad (15)$$

obeying the stationary point stability condition

$$p \left| e^h \partial_u D_{\kappa-1}(u) \right| < 1 \quad (16)$$

From (5, 12-15) we have the following expression for generating function of $\kappa$− clusters' size distribution in HD bond percolation

$$e^{-h} G_k(e^h, p) = \tilde{G}_k(u, p)\big|_{u=u(e^h, p, \kappa)}, \quad \tilde{G}_k(u,p) = 1 + \frac{z}{2}\left\{[u_k(p) - u]D_{k-1}(u) + 2\int_p^u dx D_{k-1}(x)\right\} \quad (17)$$

To derive (17) we used the identity

$$z \int_0^u dx D_{k-1}(x) = (1-p) D_k(u) + u D_{k-1}(u) \quad (18)$$

Also it follows from Eq. (18)

$$(z-1)D_{k-1}(u) = (1-p)D'_k(u) + uD'_{k-1}(u) \tag{19}$$

$$(z-2)D'_{k-1}(u) = (1-p)D''_k(u) + uD''_{k-1}(u) \tag{20}$$

Here $D'_{\kappa-1}(u) \equiv \partial_u D_{\kappa-1}(u)$, $D''_{\kappa-1}(u) \equiv \partial_u^2 D_{\kappa-1}(u)$.

Thus we have for the average number of $\kappa-$ clusters (per site)

$$N_{cl}^{(\kappa)}(p) = 1 + \frac{r_\kappa(p)-p}{p}\left[1 + r_\kappa(p) - u_\kappa(p)\right] - \frac{z}{2p}\left[r_\kappa(p) - u_\kappa(p)\right]^2 + \binom{z-1}{\kappa-1}(1-p)^{z-\kappa+1}\left[p^{\kappa-1} - r_\kappa^{\kappa-1}(p)\right] \tag{21}$$

Here $r_\kappa(p) \equiv u(1, p, \kappa)$ is the solution to the equation

$$r_\kappa - p = p\left[D_{\kappa-1}(r_\kappa) - D_{\kappa-1}(p)\right], \tag{22}$$

obeying the condition

$$p\left|\partial_{r_\kappa} D_{\kappa-1}(r_\kappa)\right| < 1, \tag{23}$$

cf. Eqs. 15, 16.

For the fraction of sites belonging to the giant $\kappa-$ cluster

$$S_\kappa(p) = 1 - \partial_\zeta G_\kappa(\zeta, p)\big|_{\zeta=1} = 1 - \tilde{G}_k\left[r_\kappa(p), p\right] - \partial_u \tilde{G}_\kappa(u,p)/\partial_u \zeta\big|_{u=r_\kappa(p)}$$

we get using (22)

$$S_\kappa(p) = 1 - N_{cl}^{(\kappa)}(p) - \frac{z}{2p}\left[r_\kappa(p) - u_\kappa(p)\right]^2 = \frac{p - r_\kappa(p)}{p}\left[1 + r_\kappa(p) - u_\kappa(p)\right] + \binom{z-1}{\kappa-1}(1-p)^{z-\kappa+1}\left[r_\kappa^{\kappa-1}(p) - p^{\kappa-1}\right] \tag{24}$$

There is the solution $r_\kappa(p) = p$ to Eq. 22. According to (23) it is stable when

$$pD'_{\kappa-1}(p) = \sum_{m=\kappa-1}^{z-1}\binom{z-1}{m}mp^m(1-p)^{z-1-m} < 1 \tag{25}$$

In this region the percolation cluster is absent, $S_\kappa(p) = 0$, and

$$N_{cl}^{(\kappa)}(p) = 1 - \frac{z}{2p}\left[p - u_\kappa(p)\right]^2 = 1 - \frac{z}{2}pD_{\kappa-1}^2(p).$$

When condition (25) breaks, another stable solution to (22) emerges

$$r_\kappa(p) = p(1-\delta_\kappa), \quad \delta_\kappa > 0. \tag{26}$$

For small $\delta_\kappa$ we have from (19, 20, 22)

$$3p^2 D''_{\kappa-1}(p)\delta_\kappa \approx 6\left[pD'_{\kappa-1}(p)-1\right] + p^2 D'''_{\kappa-1}(p)\delta_\kappa^2, \tag{27}$$

while the stability condition (23) becomes

$$p^2 D''_{\kappa-1}(p)\delta_\kappa > \left[pD'_{\kappa-1}(p)-1\right]$$

justifying the stability of the solution (26). With it we get from (21, 24)

$$N_{cl}^{(\kappa)}(p) \approx 1 - \frac{z}{2}pD^2_{\kappa-1}(p) + \frac{z}{12}D''_{\kappa-1}(p)p^3\delta_\kappa^3 \tag{28}$$

$$S_\kappa(p) \approx zpD^2_{\kappa-1}(p)\delta_\kappa \tag{29}$$

Thus at concentration $p_\kappa$ defined through the equation

$$p_\kappa D'_{\kappa-1}(p_\kappa) = \sum_{m=\kappa-1}^{z-1}\binom{z-1}{m}mp_\kappa^m(1-p_\kappa)^{z-1-m} = 1, \tag{30}$$

the phase transition into percolating phase takes place. As $pD'_{\kappa-1}(p) > pD'_\kappa(p)$ it follows from (30) that

$$p_{\kappa+1} > p_\kappa.$$

Near $p_\kappa$ we have from (27)

$$\delta_\kappa \approx A(p_\kappa)(p-p_\kappa), \qquad A(p_\kappa) > 0$$

Hence, at $p > p_\kappa$ the order parameter $S_\kappa \sim p - p_\kappa$ and singular part of the "thermodynamic potential" $N_{cl}^{(\kappa)}(p)$ is proportional to $(p-p_\kappa)^3$. This means that for all $\kappa$ the scaling indexes for the transition coincide with those for the ordinary percolation on Bethe lattice

$$\alpha = -1, \quad \beta = 1. \tag{31}$$

The Eq. (30) was first obtained in Refs. [7, 8] for HD site percolation on Bethe lattice so critical concentrations for HD site and bond percolation on it are the same. This is the consequence of strict relation between number of sites and bonds $b$ of Bethe clusters $b = s-1$ and equal numbers of empty sites and edges in their perimeters. Also $p_0 = p_1 = p_2 = (z-1)^{-1}$ for apparent reasons [7, 8].

From (17) we can also obtain the explicit expression for the size distribution of $\kappa$ - clusters $v_s^{(\kappa)}(p)$ using the Lagrange inversion formula [19] for the implicitly defined expansion

$$\tilde{G}_k(u(\zeta, p, \kappa), p) = \sum_{s=0} v_{s+1}^{(k)}(p)\zeta^s, \quad \zeta \equiv e^h \tag{32}$$

From (15, 17, 32) we get for $s > 1$

$$v_s^{(\kappa)}(p) = \frac{p^{s-1}}{(s-1)!}\partial_u^{s-2}\left[D^{s-1}_{\kappa-1}(u)\partial_u\tilde{G}_\kappa(u,p)\right]\Big|_{u=u_\kappa(p)} = z\frac{p^{s-1}}{s!}\partial_u^{s-2}D^s_{\kappa-1}(u)\Big|_{u=u_\kappa(p)}, \tag{33}$$

while from (15, 17) we get

$$v_1^{(\kappa)}(p) = 1 + z \int_p^{u_\kappa(p)} dx D_{\kappa-1}(x) = \sum_{m=0}^{\kappa-1} \binom{z}{m} p^m (1-p)^{z-m} + \sum_{m=\kappa}^{z} \binom{z}{m} u_\kappa^m(p)(1-p)^{z-m} \quad (34)$$

Using the integral representation of Eq. (33)

$$v_s^{(\kappa)}(p) = \frac{z}{s(s-1)} \oint_{|u-u_\kappa(p)|=1} \frac{du}{2\pi i} \left[\frac{pD_{\kappa-1}(u)}{u-u_\kappa(p)}\right]^{s-1} D_{\kappa-1}(u) \quad (35)$$

and the method of steepest descent we get the large $s$ asymptotic

$$v_s^{(\kappa)}(p) \sim s^{-5/2} \exp{-h_\kappa(p)s}, \quad h_\kappa(p) = \ln\frac{u^* - u_\kappa(p)}{pD_{\kappa-1}(u^*)}$$

where $u^*$ is the solution to the equation

$$D_{\kappa-1}(u^*) = D'_{\kappa-1}(u^*)\left[u^* - u_\kappa(p)\right] \quad (36)$$

For $p = p_\kappa$ we get from (22, 30, 36) $u^* = p_\kappa$, $h_\kappa(p) = 0$ and

$$\partial_p h_\kappa(p)\big|_{p=p_\kappa} = \partial_p \ln\frac{p_\kappa - u_\kappa(p)}{p}\bigg|_{p=p_\kappa} = \frac{p_\kappa D'_{\kappa-1}(p_\kappa) - 1}{p_\kappa D_{\kappa-1}(p_\kappa)} = 0$$

Hence, for $p$ close to $p_\kappa$ $h_\kappa(p) \sim (p-p_\kappa)^2$ so

$$\chi_\kappa(p) = \sum_s s^2 v_s^{(\kappa)}(p) \sim |p-p_\kappa|^{-1}.$$

Thus the critical index $\gamma = 1$ in accordance with scaling relation $\alpha + 2\beta + \gamma = 2$, see (31).

When $D_{\kappa-1}(u)$ (14) has a simple form we get simple expressions for $v_s^{(\kappa)}(p)$ for $s>1$ from Eq. (33)

$$v_s^{(0,1)}(p) = z\frac{p^{s-1}(1-p)^t}{st}\binom{s(z-1)}{s-1},$$

$$v_s^{(2)}(p) = \frac{z}{s} p^{s-1}(1-p)^t \sum_{n=0}^{s} \binom{s}{n}\binom{(s-n)(z-1)}{s-1}(-1)^n \frac{\left[1+p(1-p)^{z-2}\right]^{t-n(z-1)}}{t-n(z-1)},$$

$$v_s^{(z-1)}(p) = \frac{z}{s} p^{s-1} \sum_{n=0}^{s} \binom{s}{n}\binom{(s-n)(z-1)}{s-1}\frac{\left[(z-1)(1-p)\right]^n}{t-n(z-1)} u_{z-1}^{t-n},$$

$$u_{z-1} = p - p^z - (z-1)(1-p)p^{z-1},$$

$$v_s^{(z)}(p) = z\frac{p^{s-1}(p-p^z)^t}{st}\binom{s(z-1)}{s-1}.$$

Here $t = (z-2)s + 2$ is the perimeter of $s$-site cluster (the number of empty edges surrounding it [20]).

$v_s^{(0,1)}(p)$ coincide with that of classical bond percolation [19] as expected.

In the limit of infinite coordination number, $z \to \infty$ and $pz \to c$ ensemble of percolation patterns on the Bethe lattice becomes equivalent to random Erdos-Renyi (ER) graph [15-17]. Introducing new variable $v = zu$ we find in this limit

$$D_{\kappa-1}(v) = e^{-c} \sum_{m=\kappa-1}^{\infty} \frac{v^m}{m!} \qquad (37)$$

and from (15-17) we have

$$e^{-h}G_k(e^h, p) = \tilde{G}_k(v, p)\Big|_{v=v(e^h, p, \kappa)}, \quad \tilde{G}_k(v, p) = 1 + \frac{1}{2}\left\{[v_k(p) - v]D_{k-1}(v) + 2\int_c^v dx D_{k-1}(x)\right\}, \qquad (38)$$

where $v_\kappa(p) = ce^{-c}\sum_{m=0}^{\kappa-2}\frac{v^m}{m!}$ and $v = v(e^h, p, \kappa)$ is the solution to the equation

$$v = v_\kappa(p) + e^h c D_{\kappa-1}(v), \qquad (39)$$

which obeys the condition $c\left|e^h \partial_v D_{\kappa-1}(v)\right| < 1$.

These equations describe the infinite series of $\kappa$ - clusters percolation transitions on ER graph, which take place at critical connectivities $c_\kappa$ defined by the equation

$$c_\kappa D'_{\kappa-1}(c_\kappa) = e^{-c_\kappa}\sum_{m=\kappa-2}^{\infty}\frac{c_\kappa^{m+1}}{m!} = 1 \qquad (40)$$

In (39, 40) $\frac{1}{m!}$ is the relic of combinatorial coefficients assumed to be zero at negative $m$ so $c_0 = c_1 = c_2 = 1$ as expected for the ER graph [15-17].

As before, one can easily confirm the validity of mean-field indexes (31) for these transitions and get expression for $\kappa$ - clusters' size distribution for $s > 1$

$$v_s^{(\kappa)}(p) = \frac{c^{s-1}}{s!}\partial_v^{s-2}D_{\kappa-1}^s(v)\Big|_{v=v_\kappa(p)} \qquad (41)$$

For $\kappa = 1$ Eq. (41) gives the classical result [15-17]

$$v_s^{(1)}(p) = \frac{s^{s-2}}{s!}c^{s-1}e^{-cs}$$

and for $\kappa = 2$ we have

$$v_s^{(2)}(p) = \frac{c^{s-1}e^{-cs}}{s!}\sum_{n=0}^{s}\binom{s}{n}(-1)^{s-n}n^{s-2}e^{ce^c n}$$

For larger $\kappa$ the expressions are more cumbersome yet Eq. (41) can potentially give any number of exact relations for ER graph. Therefore, the results obtained here certify that present approach is useful and adequate.

## IV. Discussion

We can consider the present results as Bethe – Peierls approximation for Euclidean lattices with coordination number *z*. For $z \gg 1$ (say, for body-centered cubic lattice with $z = 8$) it can be quite adequate for all *p* except for a close vicinity of $p_\kappa$. To extend the region of its validity one can turn to the cluster variants of this approximation [21] in which the points of Bethe lattice are changed into unit cell of corresponding Euclidean lattice or even into the group of them. For the critical region, the real space renormalization group for $\mathcal{H}_\kappa$ in Eq. 3 can be used to obtain the approximate values of critical indexes.

The present approach can also be useful for numerical studies of HD bond percolation on real lattices. Thus, for rough estimate of clusters' size generation function the plain Metropolis Monte Carlo simulations can be used **to** obtain $Z_\kappa(h, p, q)$ for *q* = 2, 3, 4 and interpolate it to *q* = 1. To get more precise results one should extend the expression (3) for $Z_\kappa(h, p, q)$ to real *q*. This can be done, for example, within the transfer matrix representation of $Z_\kappa(h, p, q)$. This procedure is developed for 2D Potts models in Refs. [22, 23] and its application to the present model on 2D lattice is described in Appendix.

To conclude, we present here the statistical mechanics approach to HD bond percolation which is able to give the exact results for Bethe lattice, ER graph and, probably, for other hierarchical lattices. It paves the way to many analytical and numerical methods for the studies of series of HD percolation transitions in classical random bond environment on arbitrary graph.

## Acknowledgment

The work was done under financial support by the Ministry of Education and Science of the Russian Federation (state assignment Grant No. 3.5710.2017/8.9).

## APPENDIX

Here we consider the numerical procedure for description of HD percolation on 2D lattices. For *N* – column sample of 2D lattice with periodic boundary conditions partition function is expressed via transfer matrix as

$$Z_k(h, p, q) = Tr\left[\mathbf{T}^{(\kappa)}\right]^N.$$

To be specific, we consider the *L*-leg strip of square lattice. For it, we can choose the following form of the transfer matrix

$$T^{(\kappa)}_{\sigma,\sigma'}(\mathbf{n_h}, \mathbf{k}, \mathbf{n'_h}, \mathbf{k'}) = V^{(\kappa)}_\sigma(\mathbf{k}, \mathbf{n_h}, \mathbf{n'_h}) H^{(\kappa)}_{\sigma,\sigma'}(\mathbf{n'_h}, \mathbf{k}, \mathbf{k'})$$

$$V^{(\kappa)}_\sigma(\mathbf{k}, \mathbf{n_h}, \mathbf{n'_h}) = \sum_{\mathbf{n}_v} \prod_{i=1}^{L-1} \rho(n_{v,i}) g^{(\kappa)}_{\sigma_i,\sigma_{i+1}}(n_{v,i}, k_i, k_{i+1}) \prod_{i=1}^{L} \delta(k_i, n_{h,i} + n'_{h,i} + n_{v,i} + n_{v,i-1})$$

$$H^{(\kappa)}_{\sigma,\sigma'}(\mathbf{n'_h}, \mathbf{k}, \mathbf{k'}) = \prod_{i=1}^{L} \rho(n'_{h,i}) g^{(\kappa)}_{\sigma_i,\sigma'_i}(n'_{h,i}, k_i, k'_i) e^{h_{\sigma'}}$$

Here $k_i = \{0, 1, ..., 4\}$ is the coordination number of *i* – th site in the column, the same is $k'_i$ for the adjacent right column, $n_{v,i}$ is $\langle i, i+1 \rangle$ vertical edge variable in the column, $n_{h,i}$ and $n'_{h,i}$ are the horizontal edge variables for the edges joining the *i* – th site in the column from the left and from the right correspondingly.

The specific dependence of this transfer matrix on Potts variables, cf. Eqs. 1, 2, implies the limited number of distinct components in its (right) eigenvectors $v_\sigma(\mathbf{k},\mathbf{n})$ same as in ordinary $q$ - state 2D Potts model. The distinct components are characterized by the presence of sequences of equal $\sigma_i$ and distribution of zeroes among them. The scheme of numbering of distinct components is developed in context of 2D Potts model [22] and it is described in detail in Ref. [23]. Thus, eigenvectors $v_\sigma(\mathbf{k},\mathbf{n})$ can be expressed as

$$v_\sigma(\mathbf{k},\mathbf{n}) = \sum_{\alpha=1}^{d_L} R_{\sigma,\alpha} f_\alpha(\mathbf{k},\mathbf{n}),$$

where the number of distinct components $d_L$ depends on $L$ only [22, 23]. Applying the transformation to such "connectivity" basis, we get the equivalent transfer matrix

$$T^{(\kappa)}_{\alpha,\alpha'}(\mathbf{n_h},\mathbf{k},\mathbf{n'_h},\mathbf{k'}) = (R^{-1})_{\alpha,\sigma} T^{(\kappa)}_{\sigma,\sigma'}(\mathbf{n_h},\mathbf{k},\mathbf{n'_h},\mathbf{k'}) R_{\sigma',\alpha'},$$

elements of which depend on $q$ explicitly [22, 23] (as well as on $\zeta \equiv e^h$ and $p$). According to Eq. (5), at small $\zeta$ and $q$ close to 1 the largest eigenvalue of $T^{(\kappa)}_{\alpha,\alpha'}(\mathbf{n_h},\mathbf{k},\mathbf{n'_h},\mathbf{k'})$ is

$$\lambda_{max} = 1 + (q-1)G_\kappa(\zeta,p)$$

Hence, we can get $\kappa$ - clusters' size generating function $G_\kappa(\zeta,p)$ for the infinite strips of square lattice acting by $T^{(\kappa)}_{\alpha,\alpha'}(\mathbf{n_h},\mathbf{k},\mathbf{n'_h},\mathbf{k'})$ iteratively on arbitrary vector.

This scheme involves the operations with rather large matrices as $d_L \sim 5^L$ [22] and indexes $\mathbf{n_h}, \mathbf{k}$ provide also the factor $2^L \times 5^L$ to matrix dimension. However, $T^{(\kappa)}_{\alpha,\alpha'}(\mathbf{n_h},\mathbf{k},\mathbf{n'_h},\mathbf{k'})$ can be factorized into series of sparse matrices [22, 23] making the procedure amenable for computer calculations. Therefore, it may need less computer time compared to the direct enumeration of $\kappa$ - clusters on long strips in a number of bond configurations, the more so as we get in this scheme $G_\kappa(\zeta,p)$ for infinite strips. Having $G_\kappa(\zeta,p)$ for several $L$ one can get critical concentrations and critical indexes of infinite square lattice from the finite – size scaling [24].